\documentclass[preprint,12pt]{elsarticle}
\journal{Computer Physics Communications}

\usepackage{graphicx}
\usepackage{epstopdf, epsfig}
\usepackage{color}
\usepackage{booktabs}
\usepackage{bm}

\newcommand{\im}{{\rm i}}
\newcommand{\e}{{\rm e}}
\newcommand{\be}{\begin{equation}}
\newcommand{\ee}{\end{equation}}
\newcommand{\ba}{\begin{eqnarray}}
\newcommand{\ea}{\end{eqnarray}}
\newcommand{\bas}{\begin{eqnarray*}}
\newcommand{\eas}{\end{eqnarray*}}
\newcommand{\p}{\partial}
\newcommand{\vphi}{\varphi}
\newcommand{\Ah}{A_{\|}^{\rm(h)}}
\newcommand{\As}{A_{\|}^{\rm(s)}}
\newcommand{\wt}[1]{\widetilde{#1}}

\newcommand{\fe}[1]{\wt{#1}}

\newcommand{\vc}[1]{
{\bm{#1}}
}

\newcommand{\intl}{
\int \limits
}

\newcommand{\pard}[2]{
\frac{\partial #1}{\partial #2}
}

\newcommand{\df}{
{\rm d}
}

\newcommand{\totd}[2]{
\frac{\df #1}{\df #2}
}

\newcommand{\veps}{\varepsilon}

\newcommand{\gav}[1]{
\langle #1 \rangle
}
\newcommand{\Bgav}[1]{
\Big\langle #1 \Big\rangle
}
\begin{document}

\begin{frontmatter}
\title{Pullback scheme implementation in ORB5}

\author[hgw]{A.~Mishchenko}\ead{alexey.mishchenko@ipp.mpg.de}
\author[gar]{A.~Bottino}
\author[gar]{A.~Biancalani}
\author[gar]{R.~Hatzky}
\author[gar]{T.~Hayward-Schneider}
\author[spc]{N.~Ohana}
\author[spc]{E.~Lanti}
\author[spc]{S.~Brunner}
\author[spc]{L.~Villard}
\author[hgw]{M.~Borchardt}
\author[hgw]{R.~Kleiber}
\author[hgw]{A.~K\"onies}

\address[hgw]{Max Planck Institute for Plasma Physics,
  D-17491 Greifswald, Germany}
\address[gar]{Max Planck Institute for Plasma Physics,
   D-85748 Garching, Germany}
\address[spc]{Ecol\'e Polytechnique F\'ed\'erale de Lausanne (EPFL), Swiss
 Plasma Center (SPC), CH-1015 Lausanne, Switzerland}

\begin{abstract}
The pullback scheme is implemented in the global gyrokinetic particle-in-cell code
ORB5 [S.~Jolliet~et~al, Comp.~Phys.~Comm., {\bf 177}, 409 (2007)] to mitigate
the cancellation problem in electromagnetic simulations. The equations and the discretisation
used by the code are described. Numerical simulations of the Toroidal Alfv\'en
Eigenmodes are performed in linear and nonlinear regimes to verify the
scheme. A considerable improvement in the code efficiency is observed. For the
internal kink mode, it is shown that the pullback mitigation efficiently cures
a numerical instability which would make the simulation more costly otherwise.
\end{abstract}

%\maketitle
\end{frontmatter}

\section{Introduction}

Electromagnetic effects are important in fusion plasmas. Alfv\'en waves,
Magneto-Hydro-Dynamic (MHD) activity, electromagnetic modifications of the
drift waves and the turbulent transport are well-known examples. In many
cases, a combination of the global, electromagnetic and kinetic
contributions is essential. Such complexity usually cannot be addressed
analytically and calls for a numerical approach, certainly in realistic
magnetic geometries under realistic fusion plasma conditions.

Global gyrokinetic particle-in-cell simulations represent such an
approach. In this paper we focus on a particular code of this type, ORB5
\cite{jolliet_orb5}. This code has been intensively used for gyrokinetic
turbulence studies, usually in the electrostatic regime
\cite{mcmillan_2008}. The electromagnetic simulations have been inhibited by
the so-called cancellation problem \cite{Chen_Parker_2001,Mishchenko1}. In ORB5, this
problem has been mitigated using the control variate approach
\cite{Hatzky_2007}. This mitigation technique has been used for
electromagnetic microturbulence simulations \cite{Bottino_2011}, and for the
simulations of Toroidal Alfv\'en Eigenmodes \cite{Biancala_PoP16,Biancala_PPCF16}.
In this paper we describe an implementation in ORB5 of another
mitigation scheme, the so-called pullback mitigation
\cite{Mishchenko_MHD,Mishchenko_pullback,Kleiber_pullback,Mishchenko_PoPLee}. This approach
can be used in combination with the control variate scheme \cite{Hatzky_2007}. As a consequence,
the code efficiency improves considerably. Previously, the pullback mitigation
has been implemented in the EUTERPE code
\cite{Mishchenko_pullback,Kleiber_pullback,Mishchenko_PoPLee,MCole_PoP16,MCole_PoP18}. An
alternative approach to the cancellation mitigation has been recently proposed
for the GTC code \cite{Bao_Lin,Bau_Lin_Lu}.

In our simulations, we consider the Toroidal Alfv\'en Eigenmodes
\cite{Cheng_Chen_Chance,Zonca:review} destabilised 
by the fast particles \cite{Koenies_itpa,Koenies_itpa_NF} and the internal
kink instability \cite{Shafranov_70} in tokamak geometry. To our knowledge, this is the first time
the internal kink instability has been simulated using a global fully gyrokinetic
particle-in-cell code in tokamak geometry at a realistic value of
plasma~$\beta$. For simulations in straight tokamak,
%at small $\beta < m_e/m_i$,
see Refs.~\cite{SydoraNaitou95,Naitou2009,Mishchenko_kink}.

The paper is organised as follows. In Sec.~\ref{theory}, the equations solved
by ORB5 are presented. In Sec.~\ref{Discretisation}, the discretisation used
by the code is discussed. Simulations using the newly implemented schemes are
presented in Sec.~\ref{Simulations}. Conclusions are made in Sec.~\ref{conclusions}.

\section{Equations solved by ORB5}  \label{theory}
The global gyrokinetic particle-in-cell code ORB5
\cite{jolliet_orb5} solves the  gyrokinetic Vlasov-Maxwell system of
equations~\cite{Brizard:review}. The species distribution function $f_s$ is
split into the ``background'' control variate $F_{0s}$ and the
time-dependent deviation from the control
variate $\delta f_s$ so that $f_s = F_{0s} + \delta f_s$. Here, the subscript $s = {\rm i,e,f}$
indicates the particle species (bulk plasma ions and
electrons, fast particles). The control variate is usually chosen to be a
Maxwellian. The deviation from the control variate $\delta f_s$ is found from the
gyrokinetic Vlasov equation:
\be
\label{vlasov}
\pard{\delta f_{s}}{t} + \dot{\vc{R}} \cdot \pard{\delta f_{s}}{\vc{R}}\Big|_{v_{\|}} +
\dot{v}_{\|} \pard{\delta f_{s}}{v_{\|}} =
{}- \dot{\vc{R}}^{(1)} \cdot \pard{F_{0s}}{\vc{R}}\Big|_{\veps} -
\dot{\veps}^{(1)} \pard{F_{0s}}{\veps}
\ee
Here, $[\dot{\vc{R}}, \dot{v}_{\|}]$ correspond to the
gyrocenter trajectories with $[\dot{\vc{R}}^{(1)}, \dot{\veps}_{\|}^{(1)}]$ the perturbations
of the trajectories proportional to the field fluctuations. Note that the
spatial derivative at the right hand side of Eq.~(\ref{vlasov}) is taken at a constant energy $\veps
= v_{\|}^2/2 + \mu B$ whereas the spatial derivative on the left hand side of
Eq.~(\ref{vlasov}) is taken at a constant parallel velocity $v_{\|}$. Here,
$\mu = v_{\perp}^2/(2 B)$ is the magnetic moment.
The mixed-variable \cite{Mishchenko_pullback} perturbed equations of motion are
\ba
\label{dotR1}
\dot{\vc{R}}^{(1)} &=& \frac{\vc{b}}{B_{\|}^*} \times \nabla \Bgav{ \phi -
  v_{\|} A_{\|}^{\rm(s)} - v_{\|} A_{\|}^{\rm(h)} } -
\frac{q_s}{m_s} \,\gav{A^{\rm(h)}_{\|}} \, \vc{b}^* \\
\label{dotp1}
\dot{v}_{\|}^{(1)} &=& \,-\, \frac{q_s}{m_s} \,
\left[ \vc{b}^* \cdot \nabla \Bgav{\phi - v_{\|} A_{\|}^{\rm(h)}} + \pard{}{t}
  \Bgav{A_{\|}^{\rm(s)}} \right] \nonumber \\
&&{} \,-\,  \mu \, \frac{\vc{b} \times \nabla B}{B_{\|}^*} \cdot \nabla
  \Bgav{A_{\|}^{\rm(s)}}
\ea
Here, $\phi$ is the perturbed electrostatic potential, $\Ah$ and $\As$ are the
Hamiltonian and the symplectic parts \cite{Mishchenko_pullback} of the perturbed
magnetic potential, $m_s$ is the mass of the particle,
%The equilibrium magnetic field enters through the quantities
$B_{\|}^{*} = \vc{b} \cdot \nabla \times \vc{A}^{*}$,
$\vc{b}^{*} = \nabla \times \vc{A}^{*} / B_{\|}^{*}$,
$\vc{A}^{*} = \vc{A} + (m_s v_{\|}/q_s) \vc{b}$
is the modified vector potential, $\vc{A}$ is the magnetic
potential corresponding to the equilibrium magnetic field, $\vc{B} = \nabla
\times \vc{A}$, $\vc{b} = \vc{B}/B$ is the unit vector in the direction of the
equilibrium magnetic field. The gyro-averaged potential is defined as usual
$\gav{\phi} = \oint \phi (\vc{R} + \vc{\rho}) \df \alpha / (2 \pi)$
with $\vc{\rho}$ the gyroradius of the particle and $\alpha$ the gyro-phase.
Note that some nonlinear terms \cite{Kleiber_pullback} are not included in
Eqs.~(\ref{dotR1}) and (\ref{dotp1}). These terms will be considered in Sec.~\ref{nlin_pullback}.

The perturbed electrostatic potential is found from the gyrokinetic quasineutrality equation:
\be
\label{qasi}
{} - \nabla \cdot \left[ \left(\sum_{s = {\rm i,f}} \frac{q_s^2 n_{s}}{T_{s}} \rho_{s}^{2}\right)
  \nabla_{\perp} \phi \right]
= \sum_{s = {\rm i,e,f}} q_s n_{1s}
\ee
where $n_{1s} = \int \df^{6}Z \, \delta f_{s} \, \delta(\vc{R} +
\bm{\rho} - \vc{x})$ is the perturbed gyrocenter density,
%(distinct from the physical current),
$\rho_{s} = \sqrt{m_{s}T_{s}}/(q_s B)$ is the thermal gyroradius, $q_s$ is the
charge of the particle, and $\df^{6}Z = B_{\|}^{*} \, \df \vc{R} \, \df v_{\|} \,
\df \mu \, \df \alpha$ is the phase-space volume.
%corresponding to a particular species.
The polarization density is treated in
the long-wavelength approximation and finite Larmor radius
(FLR) effects are neglected for electrons. The zeroth-order densities
of the particle species satisfy the quasineutrality equation $\sum_s
q_s n_{0s} = 0$ with $s={\rm i,e,f}$.

The perturbed energy evolves according to the equation:
\ba
\dot{\veps}^{(1)} &=& v_{\|} \dot{v}_{\|}^{(1)} + \mu \dot{\vc{R}}^{(1)}
\cdot \nabla B  =
\,-\, \frac{q_s}{m_s} \left[ m_s \mu \frac{\vc{b}
    \times \nabla B}{q_s B_{\|}^*} +
  \frac{m_s v_{\|}^2}{q_s B_{\|}^*} \, (\nabla \times \vc{b}) \right]
\cdot \nabla \gav{\phi}  \,+\,  \nonumber \\
&&{} \frac{q_s}{m_s} v_{\|} \left[
   v_{\|}\vc{b} + m_s \mu \frac{\vc{b} \times \nabla B}{q_s B_{\|}^*} +
 \frac{m_s v_{\|}^2}{q_s B_{\|}^*} \, (\nabla \times \vc{b}) \right]
\cdot \nabla \gav{A_{\|}^{\rm(h)}} +  \\
&&{} \frac{q_s}{m_s} \mu B \left[ \nabla\cdot\vc{b} - \frac{m v_{\|}}{q B_{\|}^*}
  \frac{\nabla \times \vc{B}}{B^2} \cdot \nabla B \right] \Bgav{A_{\|}^{\rm(h)}} \nonumber
\ea
The symplectic part $\As$ of the perturbed magnetic potential is found
\cite{Mishchenko_pullback} from the equation:
\be
\label{Ohm}
\pard{}{t}A_{\|}^{\rm(s)} + \vc{b} \cdot \nabla \phi = 0
\ee
For the Hamiltonian part $\Ah$, the mixed-variable parallel Ampere's law is solved:
\be
\label{amp}
\left( \sum_{s = {\rm i,e,f}} \frac{\beta_s}{\rho_s^2} -
\nabla_{\perp}^{2} \right) A_{\|}^{\rm(h)}
= \mu_{0} \sum_{s = {\rm i,e,f}} j_{\|1s} +
\nabla_{\perp}^2 A_{\|}^{\rm(s)}
\ee
with $j_{\|1s} = \int \df^{6}Z \, v_{\|} \delta f_{s} \, \delta(\vc{R} +
\bm{\rho} - \vc{x})$ the perturbed parallel gyrocenter current.

The equations are solved employing the mixed-variable pullback algorithm
\cite{Mishchenko_pullback}:
\begin{enumerate}
\item At the end of each time step, redefine the magnetic potential splitting,
  so that the entire instantaneous value of the parallel magnetic potential
  $A_{\|}(t_i)$ is collected in its `symplectic part':
\be
\label{ic_A}
A_{\|\rm(new)}^{\rm(s)}(t_i) = A_{\|}(t_i) = A_{\|\rm(old)}^{\rm(s)}(t_i) +
A_{\|\rm(old)}^{\rm(h)}(t_i)
%\longrightarrow %\ , \;\;\; A_{\|}^{\rm(h)}(t_i) \longrightarrow 0
\ee
\item As a consequence of the new splitting, Eq.~(\ref{ic_A}), the
  `Hamiltonian' part of the vector potential must be corrected:
\be
\label{Ah_new}
A_{\|\rm(new)}^{\rm(h)}(t_i) = 0
\ee
\item For this modified splitting, the new mixed-variable distribution
  function must coincide with its symplectic-formulation counterpart. The
  symplectic-formulation distribution function is independent on the way of
  splitting and can be found invoking the
  pullback and using the old values of the
  mixed-variable distribution function and the `Hamiltonian' part of the
  parallel vector potential found solving, respectively, the gyrokinetic
  equation and Ampere's law, Eq.~(\ref{amp}), at the current time
  step $t_i$:
\be
\label{ic_f}
\delta f_{s\rm(new)}^{\rm(m)}(t_i) = \delta f_{s}^{\rm(s)}(t_i) =
\delta f_{s\rm(old)}^{\rm(m)}(t_i) + \frac{q_s \, \gav{A_{\|\rm(old)}^{\rm(h)}(t_i)}}{m_s}  \,
\pard{F_{0s}}{v_{\|}}
%= f_{1s} - \frac{q_s}{T_s} \, \Big[ v_{\|} - u_{\|s0}
%\Big] \gav{A_{\|}^{\rm(h)}} \, F_{0s}
\ee
Note here that Eq.~(\ref{ic_f}) corresponds to the linearised version of the
pullback transformation. In the fully nonlinear case \cite{Kleiber_pullback},
the orbits of the markers must be transformed with the marker weights kept
fixed during the transformation. This approach will be considered in Sec.~\ref{nlin_pullback}.
\item Proceed, explicitly solving the mixed-variable system of
  Eqs.~(\ref{dotR1})--(\ref{amp}) at the next time step $t_i + \Delta t$ in a
  usual way, but using Eqs.~(\ref{ic_A})--(\ref{ic_f}) as the initial conditions.
%
%\item At the end of the time step, repeat the transformation
%  Eqs.~(\ref{ic_A})-(\ref{Ah_new}).
\end{enumerate}
This algorithm combined with the usual control variate \cite{Hatzky_2007} is
the key technique mitigating the cancellation problem~\cite{Chen_Parker_2001}
in the electromagnetic gyrokinetic simulations using ORB5.

It is important to check the energy conservation in simulations. This may be a
complication since, in ORB5, the energy is monitored in the Hamiltonian
($p_{\|}$) coordinates whereas the gyrocenter markers are pushed in the
mixed-variable phase space. The evolution of the kinetic energy and the
definition of the field energy depend on the phase-space coordinates used to
define the gyrocenters since this definition includes field terms. To overcome
this complication in ORB5, we transform the mixed-variable distribution
function into the Hamiltonian coordinates:
\be
\label{fm_h}
% part.F90, L. 3294
\delta f_{s}^{\rm(h)}(t_i) =
\delta f_{s}^{\rm(m)}(t_i) - \frac{q_s \, \gav{A_{\|}^{\rm(s)}(t_i)}}{m_s}  \,
\pard{F_{0s}}{v_{\|}}
\ee
We can use then $\delta f_{s}^{\rm(h)}$ to compute the particle energy
transfer in the usual way \cite{bottino_sonnendruecker_2015,Tronko_2016}. For
the electromagnetic energy, we transform the mixed-variable perturbed parallel
current into the Hamiltonian-variable parallel current:
\be
\label{jm_h}
% fields.F90, L. 1119
\mu_0 j_{\|1\e}^{\rm(h)} = \mu_0 j_{\|1\e}^{\rm(m)} + \frac{\beta_{\e}}{\rho_{\e}^2} \As
\ee
In this paper, we transform only the electron perturbed current since the
corresponding correction for ions is much smaller, implying $j_{\|1i}^{\rm(h)}
\approx j_{\|1i}^{\rm(m)}$. The perturbed gyrocenter density does
not need to be transformed, $n_{\|1s}^{\rm(h)} = n_{\|1s}^{\rm(m)}$, if
the background distribution function is a Maxwellian.
The perturbed field energy is then defined in the usual way:
\be
% ami PhD thesis, p. 52, Eq. (5.37)
W_{\rm(field)} = \frac{1}{2} \; \sum_{s=\im,\e}\left(n_{1s}^{\rm(h)} \phi -
  j_{\|1s}^{\rm(h)} A_{\|} \right) \ , \;\; A_{\|} = \Ah + \As
\ee
These transformations of the distribution function, Eq.~(\ref{fm_h}) and of
the current, Eq.~(\ref{jm_h}), are applied only for the energy diagnostics and
do not disturb the simulation itself.
%
%=================================================================
%
\section{Discretisation} \label{Discretisation}
The deviation of the distribution function $f_s$ from the control variate
$F_{0s}$ is discretised in the mixed variable
\cite{Mishchenko_MHD,Mishchenko_pullback} with markers. This discretisation
can formally be written as
\be
\label{pic0}
\delta f^{\rm(m)}_{s}(\vc{R},v_{\|},\mu,t) = \sum_{\nu=1}^{N_{\rm p}}
w_{s\nu}(t)  %\frac{w_{s\nu}(t)}{2\pi J}
\delta(\vc{R} - \vc{R}_{\nu}) \delta(v_{\|} - v_{\nu \|})
\delta(\mu - \mu_{\nu}) \ ,
\ee
where $N_{\rm p}$ is the number of markers, $(\vc{R}_{\nu},v_{\nu \|},\mu_{\nu})$
are the marker phase space coordinates and $w_{s\nu}$ is the weight of a
marker. The markers move along the gyrocenter orbits. The evolution of the
marker weights $w_{s\nu}(t)$ is given by the gyrokinetic equation
(\ref{vlasov}). An alternative to this is to use the Lagrange invariance of the
full distribution function~$f_s$ along the gyrocenter orbits
\cite{jolliet_orb5}. The pullback step in the algorithm
described above modifies the marker weights according to Eq.~(\ref{ic_f}) at
the end of each time steps. The marker positions do not change during this
operation.

The perturbed potentials are discretized with the
finite-element method
\ba
&&{} \phi(\vc{x},t) = \sum_{\fe{L}}^{N_{\rm FE}} \phi_{\fe{L}}(t) \fe{\Lambda}_{\fe{L}}(\vc{x}) \\
&&{} \As(\vc{x},t) = \sum_{\fe{L}}^{N_{\rm FE}} a^{\rm(s)}_{\fe{L}}(t)
\fe{\Lambda}_{\fe{L}}(\vc{x}) \ , \;\;
\Ah(\vc{x},t) = \sum_{\fe{L}}^{N_{\rm FE}} a^{\rm(h)}_{\fe{L}}(t) \fe{\Lambda}_{\fe{L}}(\vc{x})
\ea
with $\fe{\Lambda}_{\fe{L}}(\vc{x})$ the finite elements (tensor product of B
splines, see below); $N_{\rm FE}$ the total number of the finite elements; $\phi_{\fe{L}}$,
$a^{\rm(s)}_{\fe{L}}$ and $a^{\rm(h)}_{\fe{L}}$ the spline coefficients. In
the weak formulation, the field equations
(\ref{qasi}), (\ref{Ohm}) and (\ref{amp}) become the linear algebra equations
in this discretisation with the differential operators represented by the
matrices. Thus, Eq.~(\ref{Ohm}) for $\As$ becomes
\be
\label{Ohm_fe}
\sum_{\fe{L}}^{N_{\rm FE}} B_{\fe{K}\fe{L}} \totd{a^{\rm(s)}_{\fe{L}}}{t} =
\sum_{\fe{R}}^{N_{\rm FE}} M_{\fe{K}\fe{R}}^{\rm(I)} \phi_{\fe{R}}
\ee
with the mass matrix $B_{\fe{K}\fe{L}}$ and the matrix $M_{\fe{K}\fe{L}}^{\rm(I)}$ representing
the parallel derivative:
\be
B_{\fe{K}\fe{L}} = \int \df^3 x \fe{\Lambda}_{\fe{K}} \fe{\Lambda}_{\fe{L}} \ , \;\;
M_{\fe{K}\fe{L}}^{\rm(I)} = \,-\,\int \df^3 x \fe{\Lambda}_{\fe{K}} \vc{b}
\cdot \nabla \fe{\Lambda}_{\fe{L}}
\ee
where $\df^3 x = J \,\df s \, \df \theta \, \df \vphi$, $J(s,\theta)$ is the real-space
Jacobian, $s = \sqrt{\psi/\psi_{\rm a}}$ is the flux label, $\psi$ is the poloidal
magnetic flux, $\psi_{\rm a}$ is the poloidal magnetic flux at the plasma edge,
$\theta$ is the poloidal angle, and $\vphi$ is the toroidal angle.

Similarly, the term $\nabla_{\perp}^2 \As$ appearing on the right hand side of
the parallel Ampere's law, Eq.~(\ref{amp}), is discretised in the
finite-element representation with a matrix as follows:
\ba
&&{} \nabla_{\perp}^2 \As \longrightarrow \sum_{\fe{L}}^{N_{\rm FE}}
M_{\fe{K}\fe{L}}^{\rm(II)} a^{\rm(s)}_{\fe{l}}(t) \\
&&{} M_{\fe{K}\fe{L}}^{\rm(II)} = \,-\,\int_\Omega \df^3 x \nabla_{\perp}\fe{\Lambda}_{\fe{K}}
\cdot \nabla_{\perp} \fe{\Lambda}_{\fe{L}} \,+\,\int_{\partial\Omega} \df^2\sigma\cdot \fe{\Lambda}_{\fe{K}}
\nabla_{\perp} \fe{\Lambda}_{\fe{L}}
\ea
where the second term is a boundary integral coming from integration by parts.
This trick allows to use linear finite elements, whose second derivatives are
singular.
For this Laplace matrix, the poloidal-plane approximation for $\nabla_{\perp}$
can be used. In this case, it can be treated in the same way as the Laplacian
appearing on the left hand side of Ampere's law, Eq.~(\ref{amp}), with the
only difference that $M_{\fe{K}\fe{L}}^{\rm(II)}$ is used for the
multiplication and does not need, therefore, to be inverted.

The treatment of the parallel-derivative matrix $M_{\fe{k}\fe{l}}^{\rm(I)}$ is
more complicated. This matrix must be
split into two parts corresponding to the poloidal derivative and the toroidal
derivative:
\be
M_{\fe{K}\fe{L}}^{\rm(I)} = \,-\, \int \vc{b}\cdot\nabla\theta \;
\fe{\Lambda}_{\fe{K}} \pard{\fe{\Lambda}_{\fe{L}}}{\theta} \; \df^3 x -
\int \vc{b}\cdot\nabla\vphi \;
\fe{\Lambda}_{\fe{K}} \pard{\fe{\Lambda}_{\fe{L}}}{\vphi} \; \df^3 x
\ee
These two ``poloidal'' and ``toroidal'' matrices have to be treated separately
due to the toroidal Fourier transform applied in ORB5
\cite{jolliet_orb5,mcmillan_fft} to solve the field Eqs.~(\ref{qasi}), (\ref{Ohm}) and (\ref{amp}).
Recall that ORB5 solves the gyrokinetic system of equations in an axisymmetric tokamak
geometry where the equilibrium does not depend on the toroidal angle $\vphi$.

In the three-dimensional space, one defines the finite elements
$\fe{\Lambda}_{\fe{L}}$ as the tensor products of the usual B splines
$\lambda_j(x)$, typically cubic:
\be
\label{tensor_prod}
\fe{\Lambda}_{\fe{L}}(\vc{x}) = \lambda_j(s) \lambda_k(\theta) \lambda_l(\vphi)
\ee
Note that we use capital symbols and the tilde notation throughout this paper in order to
indicate the tensor-product nature of both the finite elements $\fe{\Lambda}_{\fe{L}}$
and their indexes $\fe{L}$ evident from Eq.~(\ref{tensor_prod}). The integer
indexes $j$, $k$, and $l$ of the one-dimensional B splines change from zero to
the number of the B splines used in the respective direction.
The perturbed field, for example the electrostatic potential $\phi$, can be
represented in terms of the usual B splines as follows
\be
\phi(s,\theta,\vphi) = \sum_{l'=0}^{N_{\vphi}-1} \sum_{j'k'} \phi_{j'k'l'}
\lambda_{j'}(s) \lambda_{k'}(\theta) \lambda_{l'}(\vphi)
\ee
with $N_{\vphi}$ being the number of the toroidal B splines. The spline
coefficients $\phi_{j'k'l'}$ can be Fourier transformed in the toroidal
coordinate. The Fast-Fourier-Transform is used in ORB5 \cite{mcmillan_fft}:
\be
\phi_{j'k'l'} = \sum_{n=0}^{N_{\vphi}-1} \phi_{j'k'}^{(n)} \;
\exp\left[\frac{2 \pi \im}{N_{\varphi}} \; n l' \right]
\ee
Using this representation, one can write for the toroidal derivative
\be
\label{tor_deriv}
\pard{\phi(s,\theta,\vphi)}{\vphi} = \sum_{l'=0}^{N_{\vphi}-1} \sum_{n=0}^{N_{\vphi}-1}
\exp\left[\frac{2 \pi \im}{N_{\varphi}} \; n l' \right] \, \pard{\lambda_{l'}(\vphi)}{\vphi}
\sum_{j'k'} \phi_{j'k'}^{(n)} \lambda_{j'}(s) \lambda_{k'}(\theta)
\ee
In the weak formulation, the finite-element representation of the operator $\p
/ \p \vphi$ acting on the perturbed field $\phi$ and its toroidal Fourier
transform are represented by the tensors $B_{jkl}$ and $B_{jk}^{(n)}$ as follows:
\ba
B_{jkl} &=& \int\pard{\phi(s,\theta,\vphi)}{\vphi} \lambda_j(s)
\lambda_k(\theta) \lambda_l(\vphi) \, J(s,\theta) \, \df s \, \df
\theta \, \df \vphi \\
&=&
\sum_{n=0}^{N_{\vphi}-1} B_{jk}^{(n)} \;
\exp\left[\frac{2 \pi \im}{N_{\varphi}} \; n l \right] \nonumber
\ea
Here, the indexes $j$, $k$, $l$ and $n$ are integers. Substituting
Eq.~(\ref{tor_deriv}), we can write
\ba
B_{jkl} &=& \sum_{l'=0}^{N_{\vphi}-1} \sum_{n=0}^{N_{\vphi}-1}
\exp\left[\frac{2 \pi \im}{N_{\varphi}} \; n l' \right] \,
\intl_0^{2\pi} \pard{\lambda_{l'}(\vphi)}{\vphi} \, \lambda_l(\vphi) \, \df \vphi
\\
&&{} \times \sum_{j'k'} \phi_{j'k'}^{(n)} \, \int \lambda_{j'}(s) \lambda_{k'}(\theta)
\lambda_j(s) \lambda_k(\theta)  \, J(s,\theta) \, \df s \, \df \theta \nonumber
\ea
For the Fourier coefficients $B_{jk}^{(n)}$, we obtain:
\ba
&&{} \sum_{n=0}^{N_{\vphi}-1} B_{jk}^{(n)} \; 
\exp\left[\frac{2 \pi \im}{N_{\varphi}} \; n l \right] =
\sum_{n=0}^{N_{\vphi}-1} D^{(n)}
\exp\left[\frac{2 \pi \im}{N_{\varphi}} \; n l \right] \\
&&{} \times \sum_{j'k'} \phi_{j'k'}^{(n)} \int \lambda_{j'}(s)
\lambda_{k'}(\theta) \lambda_j(s) \lambda_k(\theta) \, J(s,\theta) \, \df s \,
\df \theta
\nonumber
\ea
Thus for the individual toroidal modes, we have to compute
\be
B_{jk}^{(n)} = D^{(n)} \sum_{j'k'} \phi_{j'k'}^{(n)} \int \lambda_{j'}(s)
\lambda_{k'}(\theta) \lambda_j(s) \lambda_k(\theta) \, J(s,\theta) \, \df s \,
\df \theta
\ee
One sees that the Fourier transform of the toroidal derivative acting on the
perturbed electrostatic potential~$\phi$ is constructed from the Fourier
transform of the spline coefficients $\phi_{j'k'}^{(n)}$, the two-dimensional
mass matrix 
\be
\int \lambda_{j'}(s)
\lambda_{k'}(\theta) \lambda_j(s) \lambda_k(\theta) \, J(s,\theta)\, \df s \, \df
\theta
\ee
and the quantity $D^{(n)}$ defined by
\ba
&&{} D^{(n)} \exp\left[\frac{2 \pi \im}{N_{\varphi}} \; n l \right] =
\sum_{l'=0}^{N_{\varphi}-1} \intl_0^{2 \pi} \df \varphi \;
\totd{\lambda_{l'}(\varphi)}{\varphi} \, \lambda_{l}(\varphi) \;
\exp\left[\frac{2 \pi \im}{N_{\varphi}} \; n l' \right] \;=\; \\
&&{} \;=\; \exp\left[\frac{2 \pi \im}{N_{\varphi}} \; n l \right]
\sum_{k=-p}^{p} \intl_0^{2 \pi} \df \varphi \;
\totd{\lambda_{l+k}(\varphi)}{\varphi} \, \lambda_{l}(\varphi) \;
\exp\left[\frac{2 \pi \im}{N_{\varphi}} \; n k \right]   \nonumber
\ea
Here, $p$ is the order of the B splines.
Using the periodicity and the symmetry relations for the B splines and their derivatives:
\ba
&&{} \intl_0^{2\pi} \lambda_l(\vphi) \, \pard{\lambda_l}{\vphi} \, \df \vphi = 0 \ , \;\;
\lambda(x) = \lambda(-x) \\
&&{} \lambda'(x) = \,-\,\lambda'(-x)
\ , \;\; \lambda'(x) = \totd{\lambda(x)}{x}
\ea
and introducing the notation $\vphi = (2\pi/N_{\vphi})\;x$, we can write
\ba
D^{(n)} &=& 2 \im \sum_{k=1}^{p} \sin \left[\frac{2 \pi}{N_{\varphi}} \; n k
\right] \; \intl_0^{2 \pi} \totd{\lambda_{l+k}(\varphi)}{\varphi} \,
\lambda_{l}(\varphi) \; \df \varphi \nonumber \\
&=& 2 \im \sum_{k=1}^p
d_k \, \sin \left[\frac{2 \pi}{N_{\varphi}} \; n k \right] \;
\ea
Here, we have introduced the notations:
\be
d_k = \intl_0^{N_{\varphi}} \totd{\lambda_{l+k}(x)}{x} \, \lambda_{l}(x) \, \df x =
\sum_{j=0}^{p-k} \intl_0^1 \totd{P_{j+k}^{(p)}(x)}{x} \, P_{j}^{(p)}(x) \, \df x \ ; \;\;
k = 1, \ldots, p
\ee
with $P_j^{(p)}$ the B spline polynomials of the degree $p$ defined within a grid cell
\cite{Hoellig}. Summarising, we obtain:
\be
D^{(n)} = 2 \im \sum_{k=1}^p d_k \sin\left(\frac{2 \pi}{N_{\varphi}} \; n k \right) \ , \;\;
d_k = \sum_{j=0}^{p-k} \intl_0^1 P_{j}^{(p)}(x) \; \totd{P_{j+k}^{(p)}(x)}{x} \; \df x
\ee
For the linear B splines, $p = 1$, we obtain
\be
d_1 = \,-\,\frac{1}{2} \ , \;\;
D^{(n)} = \,-\,\im \, \sin\left(\frac{2 \pi n}{N_{\vphi}}\right)
\ee
For the quadratic ($p = 2$) B splines:
\be
d_1 = \,-\,\frac{5}{12} \ , \;\; d_2 = \,-\,\frac{1}{24} \ , \;\;
D^{(n)} = \im \sin\left(\frac{2 \pi n}{N_{\vphi}}\right) \,
\left[ \,-\,\frac{5}{6} - \frac{1}{6} \cos\left(\frac{2 \pi n}{N_{\vphi}}\right) \right]
\ee
Finally, for the cubic ($p = 3$) B splines:
\ba
&&{} d_1 = \,-\,\frac{49}{144} \ , \;\;
d_2 = \,-\,\frac{7}{90} \ , \;\;
d_3 = \,-\,\frac{1}{720} \\
&&{} D^{(n)} = \im \sin\left(\frac{2 \pi n}{N_{\vphi}}\right) \,
\left[ \,-\,\frac{61}{90}
-\frac{14}{45} \cos\left(\frac{2 \pi n}{N_{\vphi}}\right)
-\frac{1}{90} \cos^2\left(\frac{2 \pi n}{N_{\vphi}}\right)
\right]
\ea
The poloidal derivative in the finite-element toroidal Fourier representation
must be treated differently:
\be
\label{pol_deriv}
\pard{\phi(s,\theta,\vphi)}{\theta} = \sum_{l'=0}^{N_{\vphi}-1} \sum_{n=0}^{N_{\vphi}-1}
\exp\left[\frac{2 \pi \im}{N_{\varphi}} \; n l' \right] \, \lambda_{l'}(\vphi)
\sum_{j'k'} \phi_{j'k'}^{(n)} \lambda_{j'}(s) \pard{\lambda_{k'}(\theta)}{\theta}
\ee
In the weak formulation, the operator $\p / \p \theta$ acting of $\phi(\vc{x})$,
represented by the tensors $G_{jkl}$ and $G_{jk}^{(n)}$, is
\ba
G_{jkl} &=& \int\pard{\phi(s,\theta,\vphi)}{\theta} \lambda_j(s)
\lambda_k(\theta) \lambda_l(\vphi) \, \df s \, \df
\theta \, \df \vphi \nonumber \\
&=& \sum_{n=0}^{N_{\vphi}-1} G_{jk}^{(n)} \;
\exp\left[\frac{2 \pi \im}{N_{\varphi}} \; n l \right]
\ea
Performing the same manipulations, as above, we obtain
\be
G_{jk}^{(n)} = M^{(n)} \sum_{j'k'} \phi_{j'k'}^{(n)} \int \lambda_{j'}(s)
\pard{\lambda_{k'}(\theta)}{\theta} \lambda_j(s) \lambda_k(\theta) \,
J(s,\theta) \, \df s \, \df \theta
\ee
One sees that the Fourier transform of the {\it poloidal derivative} acting on the
perturbed electrostatic potential $\phi(\vc{x})$ is constructed from the Fourier
transform of the spline coefficients $\phi_{j'k'}^{(n)}$, the %two-dimensional
non-symmetric matrix
\be
\int \lambda_{j'}(s) \pard{\lambda_{k'}(\theta)}{\theta} \lambda_j(s)
\lambda_k(\theta) \, J(s,\theta)\, \df s \, \df \theta
\ee
and the new normalisation factors $M^{(n)}$ defined by
\ba
\label{Mnfin}
&&{} M^{(n)} = \frac{2\pi}{N_\varphi}\left\{2 \, \sum_{k=1}^{p} c_k
  \cos\left(\frac{2\pi}{N_\varphi}nk\right) + c_0\right\} \\
&&{} c_k=\sum_{j=0}^{p-k}\displaystyle\int_0^1 \df x P^{(p)}_j(x) P^{(p)}_{j+k}(x)
\ , \;\; k = 0, \ldots, p
\ea
Note the dependence on the toroidal mode number for the normalisation factors
$M^{(n)}$ and $D^{(n)}$. The factor $D^{(n)}$ appears in the matrices
discretising the toroidal derivative operator. The factor $M^{(n)}$ is needed
in the poloidal derivative but also in all other operators which do not involve
$\p/\p\vphi$. The peculiarity of the poloidal derivative operator is that it
leads to non-symmetric matrices. All other matrices used in ORB5 are
symmetric.

In explicit terms, the matrix acting on the electrostatic potential in the
equation for $\As$, Eq.~(\ref{Ohm_fe}), is given by
\ba
\sum_{\fe{L}}^{N_{\rm FE}} M_{\fe{K}\fe{L}}^{\rm(I)} \phi_{\fe{L}} &=&
\sum_{n=0}^{N_{\vphi}-1}
\exp\left[\frac{2 \pi \im}{N_{\varphi}} \; n l \right] \\
&&{} \times \left[ M^{(n)} \sum_{j'k'} M_{jk;j'k'}^{\rm(I;\theta)}\phi_{j'k'}^{(n)} +
D^{(n)} \sum_{j'k'} M_{jk;j'k'}^{\rm(I;\vphi)}\phi_{j'k'}^{(n)} \right] \nonumber
\ea
with the two-dimensional matrices
\ba
&&{} M_{jk;j'k'}^{\rm(I;\theta)} = \,-\, \int \vc{b}\cdot\nabla\theta \;
\lambda_{j'}(s) \pard{\lambda_{k'}(\theta)}{\theta}
\lambda_j(s) \lambda_k(\theta) \, J(s,\theta)\, \df s \, \df \theta \\
&&{} M_{jk;j'k'}^{\rm(I;\vphi)} = \,-\, \int \vc{b}\cdot\nabla\vphi \; \lambda_{j'}(s) \lambda_{k'}(\theta)
\lambda_j(s) \lambda_k(\theta) \, J(s,\theta)\, \df s \, \df \theta
\ea
Precisely these expressions have been implemented in ORB5. The field equations
are solved for each toroidal mode individually \cite{mcmillan_fft}. The matrix
routines have been modified in ORB5 to allow for the non-symmetric matrices.
%
%=================================================================
%
\section{Simulations} \label{Simulations}
\begin{figure}[h!]
\begin{center}
\includegraphics[width=0.49\textwidth]{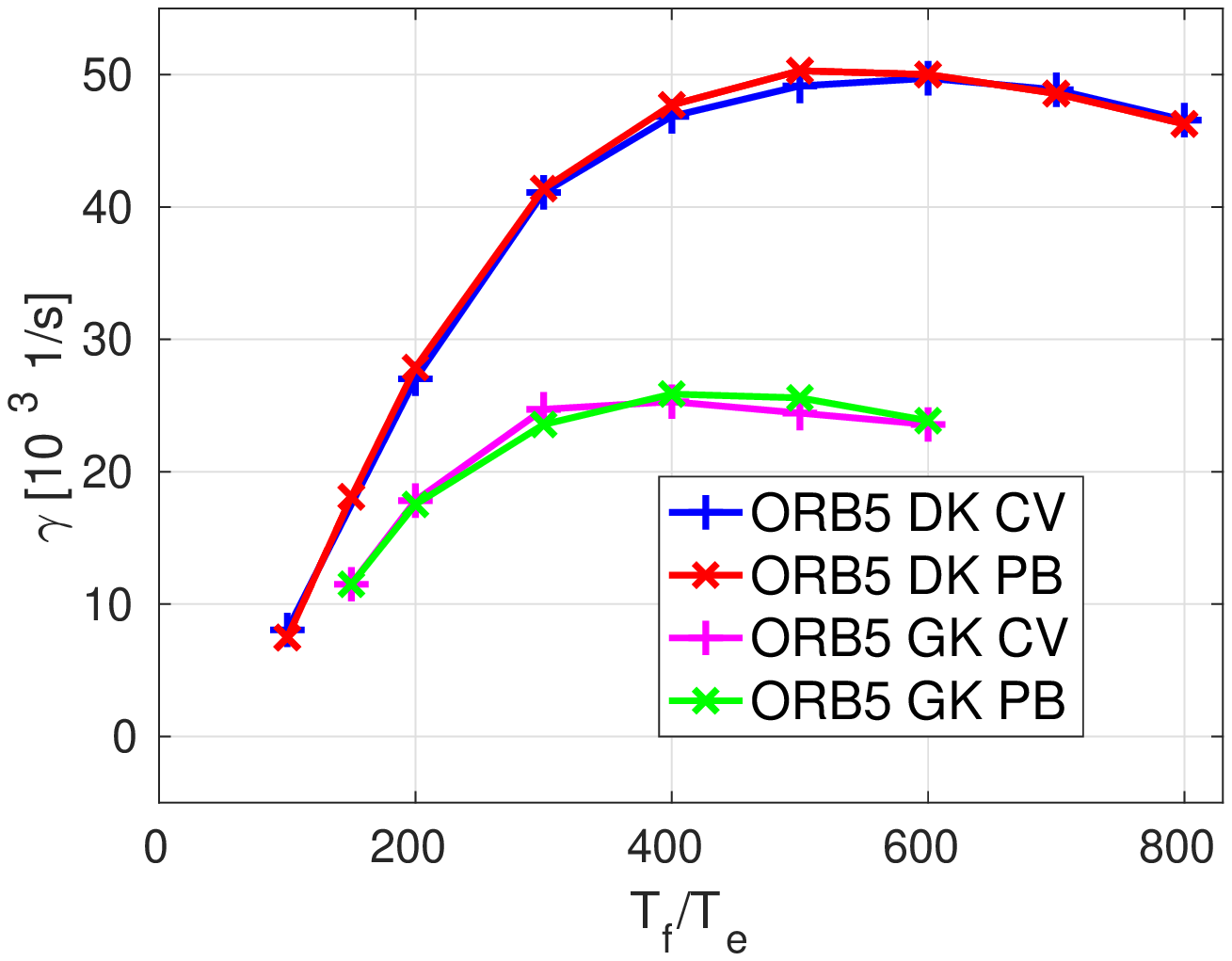}
\includegraphics[width=0.49\textwidth]{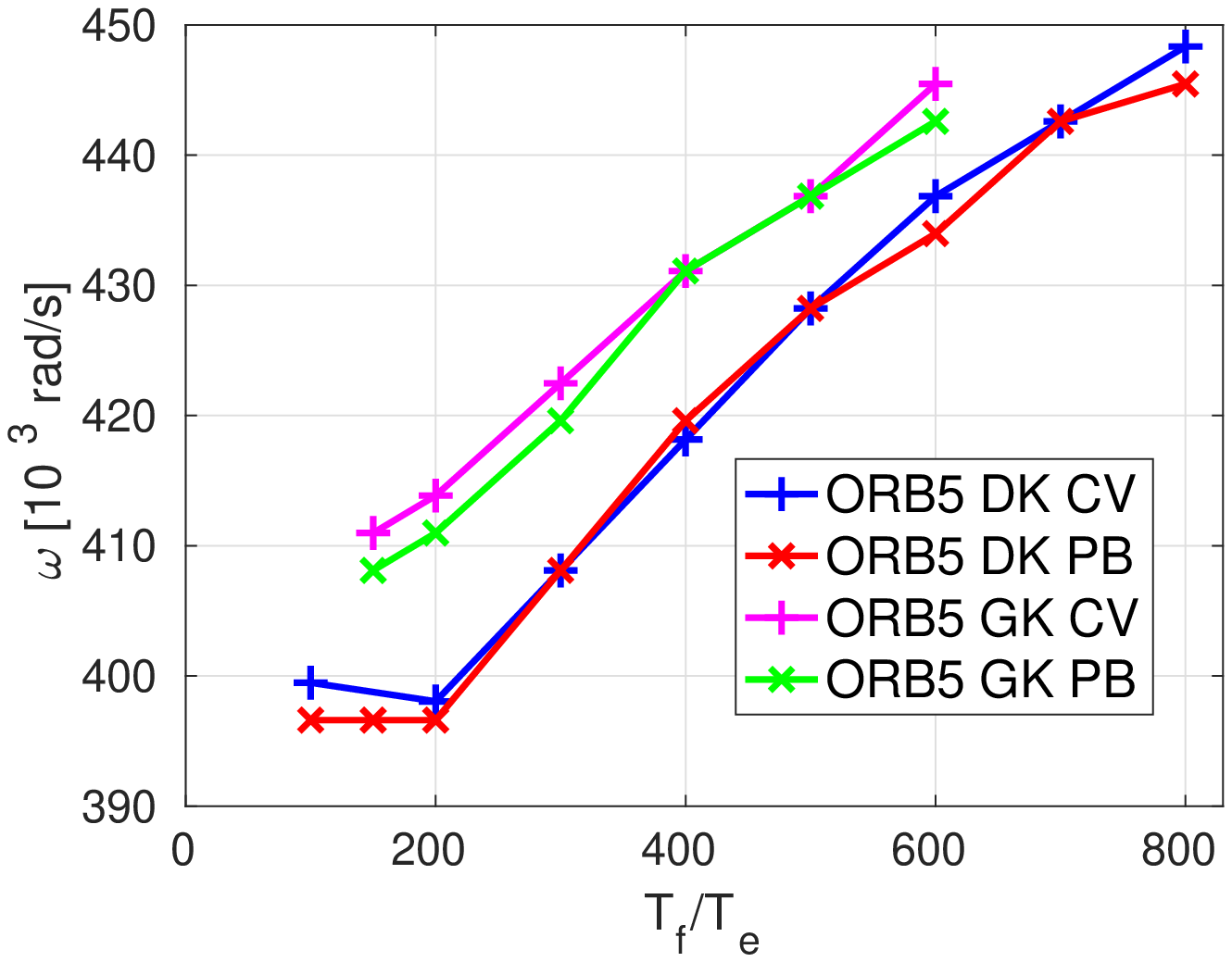}
%\vskip -1em
\caption{\label{fig:lin} Dependence of the linear growth rate and the frequency of the TAE
  instability \cite{Koenies_itpa,Koenies_itpa_NF} on the fast-ion
  temperature. Control-Variate (CV) mitigation \cite{Hatzky_2007} is compared with the pullback
  (PB) mitigation \cite{Mishchenko_pullback}. Drift-kinetic fast ions (DK) are
  compared with the gyrokinetic (GK) ones.} 
\end{center}
\end{figure}
\begin{figure}[h!]
\begin{center}
\includegraphics[width=0.8\textwidth]{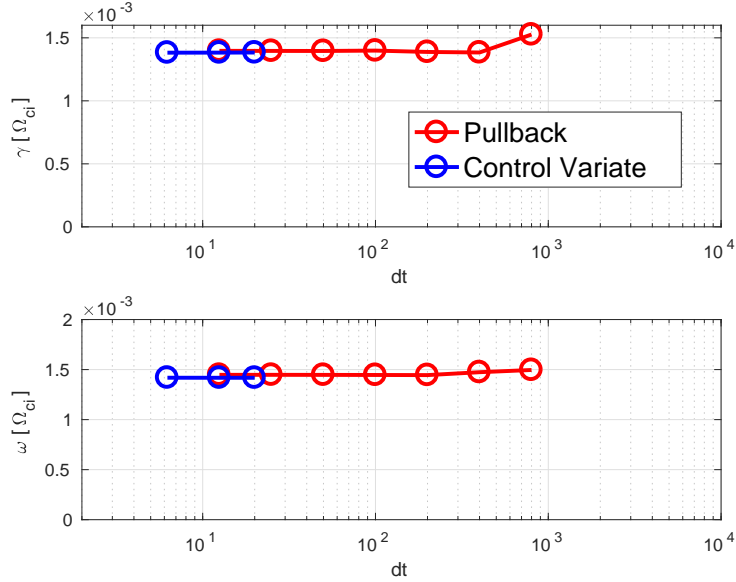}
\caption{\label{fig:tae_ami} ITPA-TAE benchmark with increased fast-ion
  density. The TAE growth rate and the frequency as functions of the time step
  used by ORB5. The time step can be considerably increased when using the
  pullback scheme compared to the control variate only, which becomes
  numerically unstable at larger time steps. The time step is measured in the
  ion-cyclotron units.}
\end{center}
\end{figure}

\begin{figure}[h!]
\begin{center}
\includegraphics[width=0.47\textwidth]{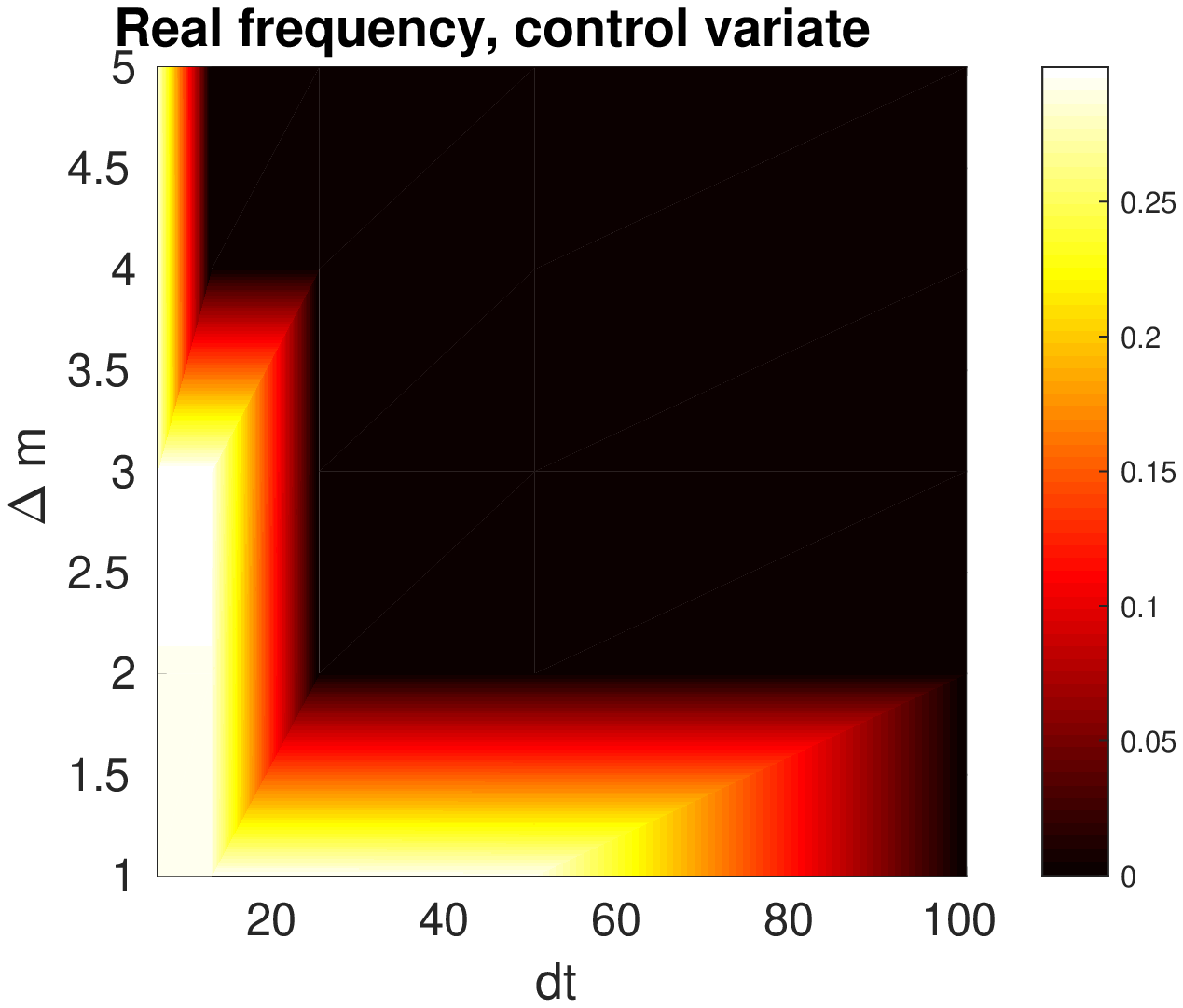}
\includegraphics[width=0.47\textwidth]{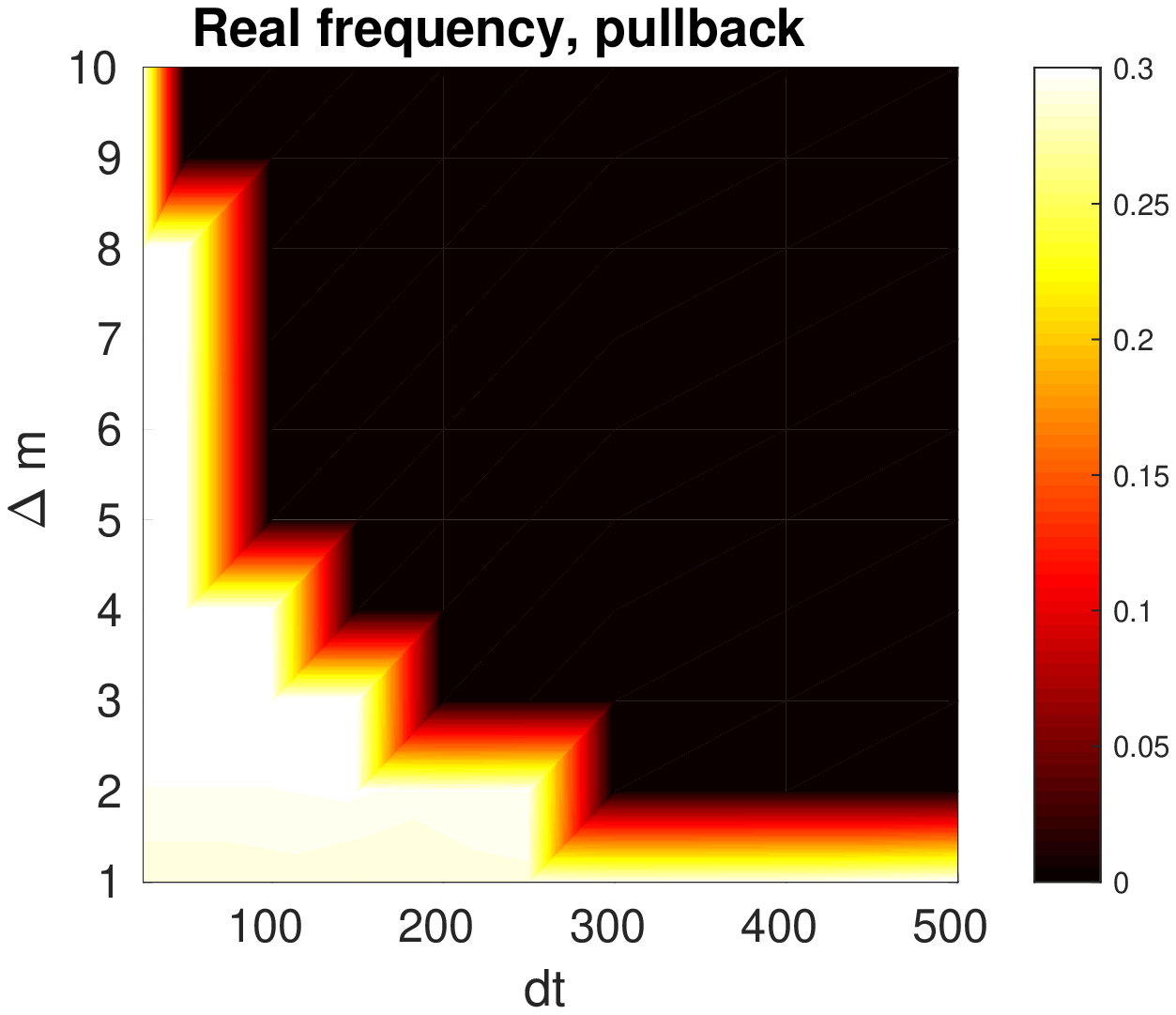}
\caption{\label{fig:omega_mat} Real frequency as a function of the diagonal
  Fourier filter width and the time step. On the left figure, only the control
  variate has been applied. The time step must be strongly reduced for
  large filters. On the right figure, the pullback scheme has been used. As
  a consequence, the time step requirements are considerably relaxed.}
\end{center}
\end{figure}

\begin{figure}[h!]
\begin{center}
\includegraphics[width=0.47\textwidth]{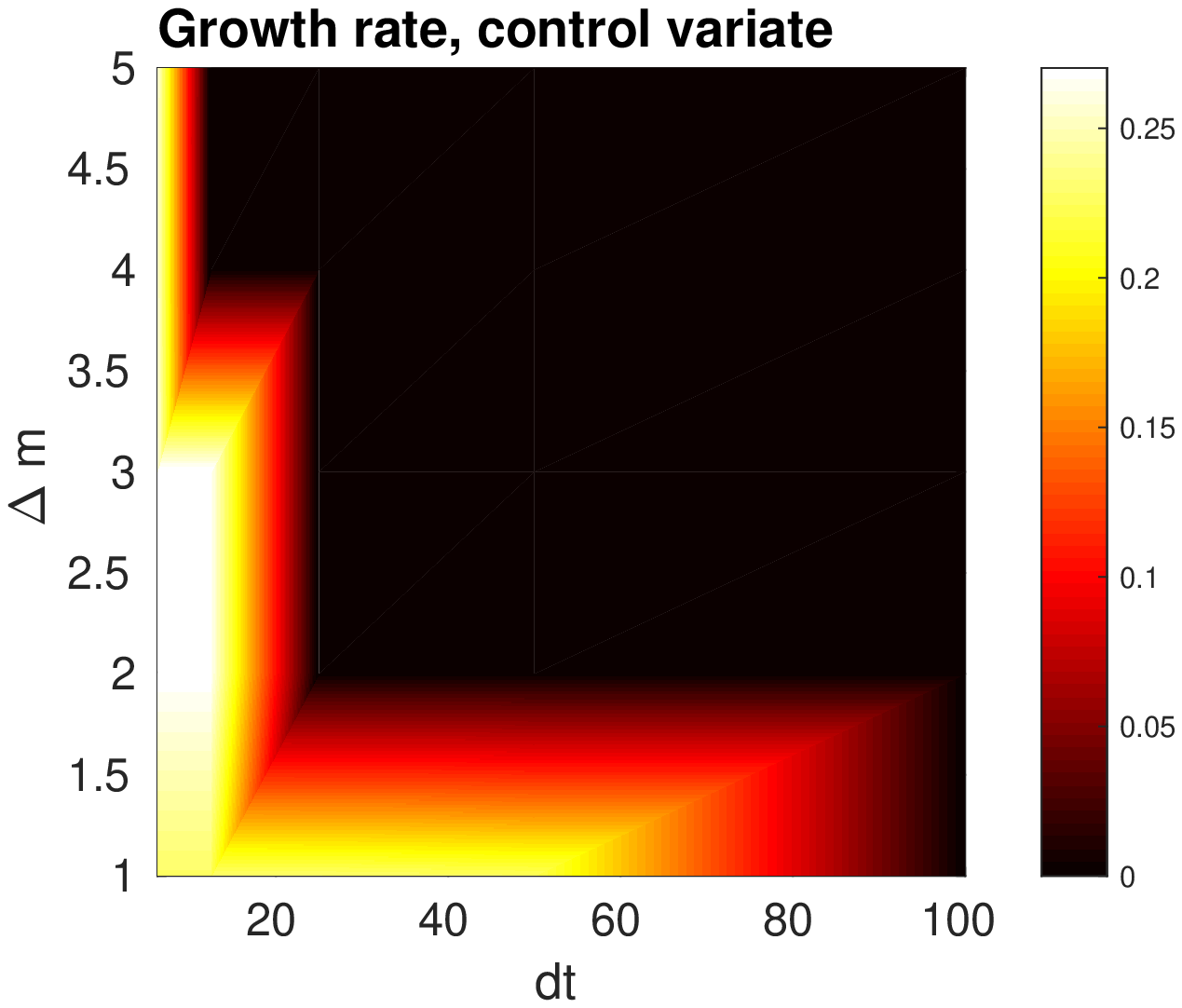}
\includegraphics[width=0.47\textwidth]{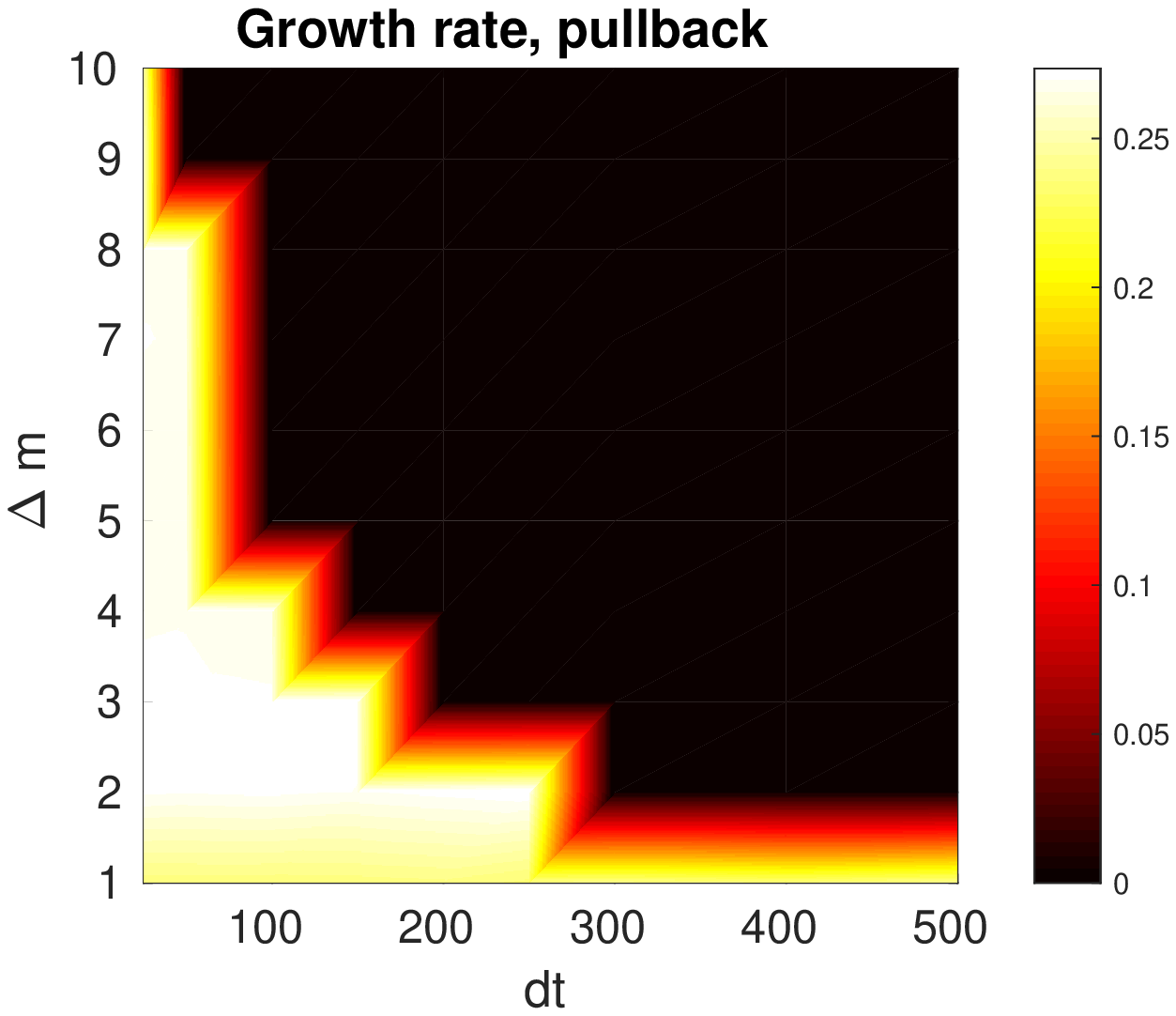}
\caption{\label{fig:gamma_mat} Growth rate as a function of the diagonal
  Fourier filter width and the time step. On the left figure, only the control variate is
  applied. One sees that the time step must be strongly reduced for large filter widths,
  similar to Fig.~\ref{fig:omega_mat}. On the right figure, the pullback
  scheme is used. The time step requirements are considerably relaxed.}
\end{center}
\end{figure}

\subsection{Toroidal Alfv\'en Eigenmode}
For verification of the scheme newly implemented in ORB5, we consider the
reference case of the international cross-code ``ITPA-TAE'' benchmark
\cite{Koenies_itpa,Koenies_itpa_NF}. In this benchmark, the Toroidal Alfv\'en
Eigenmode with the toroidal mode number $n = {}-6$ and the dominant poloidal mode
numbers $m = 10$ and $m = 11$ has been considered in the linear regime. The
mode has been studied in tokamak geometry with the small radius $r_{\rm a} =
1$~{\rm m}, the large radius $R_0 = 10$~{\rm m}, the magnetic field on the
axis $B_0 = 3$~{T}, and the safety factor profile $q(r) =  1.71 + 0.16 (r /
r_{\rm a})^2$, where $r$ is the minor radius of the plasma. The flat
background plasma profiles have been chosen with the ion density
$n_{\im} = 2 \times 10^{19}~{\rm m^{-3}}$, the ion and electron temperatures
$T_\im = T_\e = 1$~{\rm keV}, corresponding to $\beta_{\rm bulk} = 2 \mu_0 (n_{\im}
T_\im + n_{\e} T_\e)/B^2 \approx 0.18~\%$. Using the fast-ion parameters of
Refs.~\cite{Koenies_itpa,Koenies_itpa_NF}, we obtain the
result shown in Fig.~\ref{fig:lin}. Here, the mode growth rate is shown as a
function of the fast-ion temperature at the fast-ion density held constant. We
compare the simulations using the control-variate approach
\cite{Hatzky_2007} to the mitigation of the cancellation problem with the
simulations using the pullback scheme. One sees
that the agreement is very good. This verifies the pullback scheme
implementation in ORB5. 

For the control-variate simulations, we have used $N_{\e}=2 \times 10^7$ electron markers,
$N_{\im}= 10^7$ ion markers, $N_{{\rm f}} = 10^7$ fast markers, 
$N_{{\rm s}} = 256$ radial grid points, $N_{\theta} = 256$ poloidal grid
points, and $N_{\varphi} = 64$ toroidal grid points. The time step is
$\omega_{{\rm ci}} \Delta t = 20$. The Fourier filter includes the poloidal
modes $9 < m < 12$ and the toroidal mode $n=-6$. Dirichlet boundary conditions
for the potentials are set at the axis and at the edge. The ion/electron mass
ratio is $m_\im/m_\e=200$. The initial perturbation of the ion gyrocenter density is
localised near the mid-radius, and with mode numbers $m = 10,11$ and $n=\,-\,6$. 
For the simulations with the pullback mitigation, we have used the same
parameters, except the number of markers, which has been set as $N_\e=5 \times
10^6$ electron markers, $N_\im= 2 \times 10^6$ ion markers, $N_{\rm f} = 5
\times 10^6$ fast markers, and the time step, which is $\omega_{{\rm ci}} \Delta t =
100$. A numerical high-frequency instability rising at outer radii ($s \sim 0.7$) is
observed to develop in the simulations with the control variate only, and
gyrokinetic fast ions, at large times. In order to postpone the rising time of
the instability, it is necessary to increase the number of markers with
respect to the values chosen for the simulations with the pullback mitigation. 

Now, we consider the benefits of the new pullback scheme in ORB5
simulations. In Fig.~\ref{fig:tae_ami}, the growth rate and the frequency of
the TAE mode is shown as a function of the time step used by ORB5. Whereas the
control variate mitigation is capable reproducing only the first three points,
becoming numerically unstable for larger time steps, the pullback works
robustly even for time steps two orders of magnitude larger than the time
steps typically
used for the electromagnetic simulations with the control variate only. In
this simulation, the fast-ion density has been ten times larger (corresponding
to 3\% particle content) comparing to the standard
``ITPA-TAE'' benchmark \cite{Koenies_itpa,Koenies_itpa_NF}. This makes the
simulations faster and more robust.

This improvement in the numerical properties of the code can also be seen in
Figs.~\ref{fig:omega_mat} and \ref{fig:gamma_mat}. Here, we plot the frequency
and the growth rate, respectively, as functions of the numerical parameters:
the times step and the width of the diagonal Fourier filter
\cite{jolliet_orb5,mcmillan_fft} used in the ORB5 simulations. Here, we
consider the original ``ITPA-TAE'' \cite{Koenies_itpa,Koenies_itpa_NF}
parameters again. One sees that the time step needs to be adjusted depending
on the number of Fourier modes in the filter (defined by its width) both
for the control variate and for the pullback schemes. This requirement is,
however, considerably relaxed when the pullback mitigation is used.

% \begin{figure}[h!]
% \begin{center}
% \includegraphics[width=0.47\textwidth]{graph/NL-phimax_t-DKf.eps}
% \caption{\label{fig:nonlin} Nonlinear simulations with drift-kinetic fast
%   ions. The same colours are used as in Fig.~\ref{fig:lin}. Note that, not
%   only the linear growth rate, but also the nonlinear saturation level agrees
%   well with and without the pullback scheme.}
% \end{center}
% \end{figure}

% Finally, we compare the pullback scheme with the usual control variate
% approach in the nonlinear regime with only the wave-particle nonlinearity
% kept. In Fig.~\ref{fig:nonlin}, one sees that the saturation levels agree well
% for both schemes. This verifies the new pullback implementation in ORB5
% also nonlinearly. This case has been considered in more detail using the 
% control-variate mitigation in Ref.~\cite{MCole_PoP16}.

%
%~~~~~~~~~~~~~~~~~~~~~~~~~~~~~~~~~~~~~~~~~~~~~~~~~~~~~~~~~~~~~~~~~
%
\subsection{Internal kink instability}
\begin{figure}[h!]
\begin{center}
\includegraphics[width=0.47\textwidth]{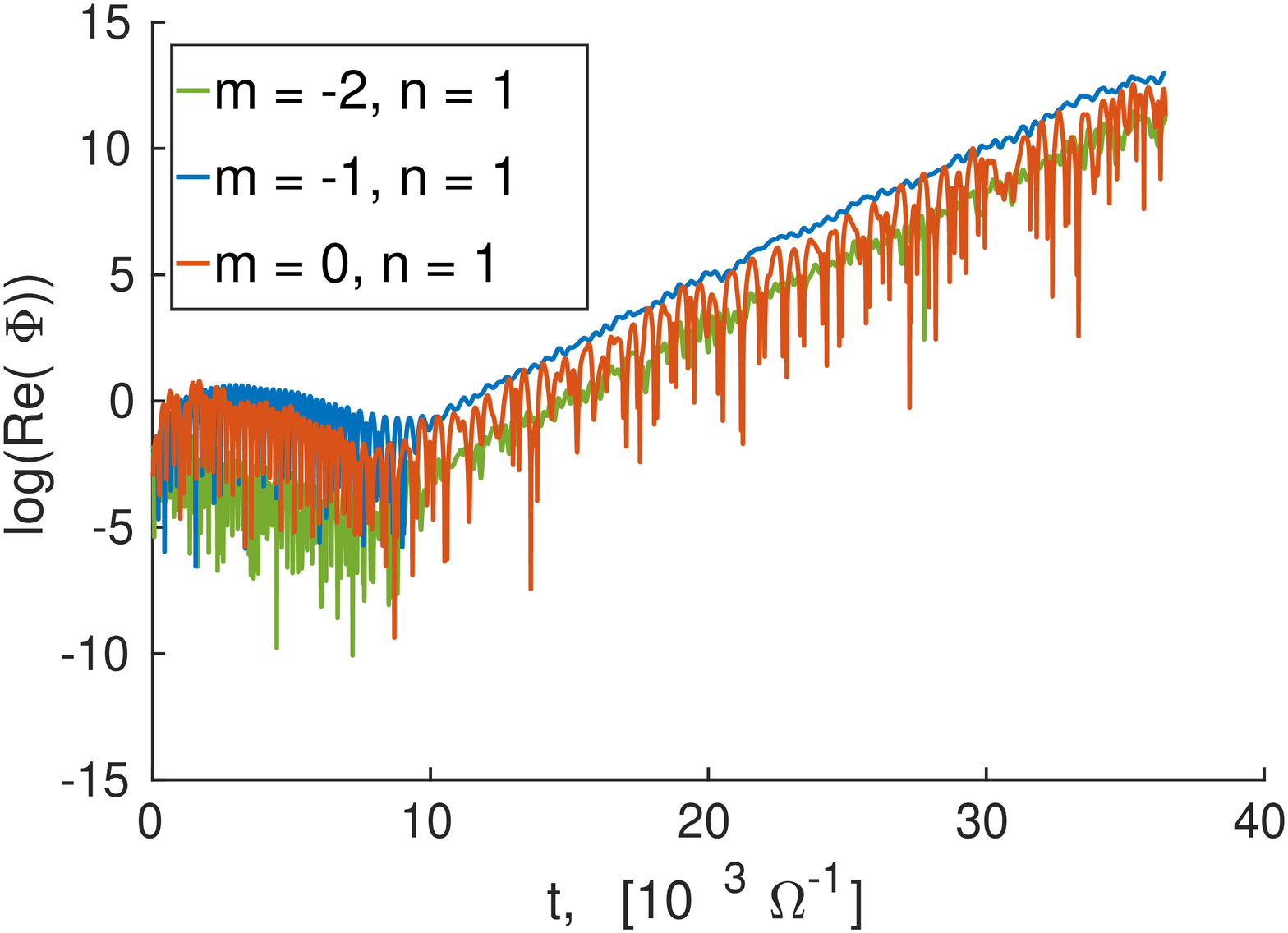}
\includegraphics[width=0.47\textwidth]{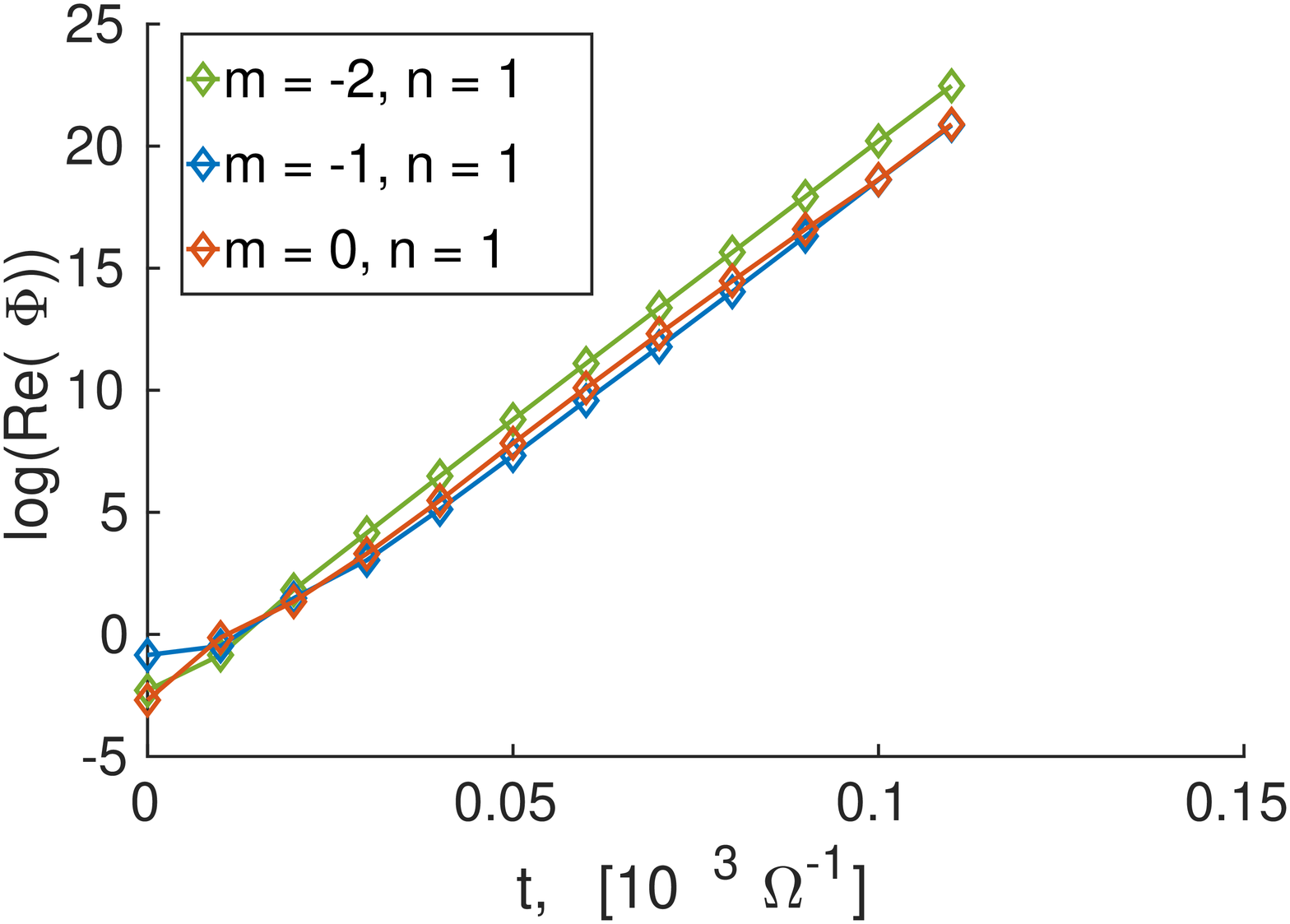}
\caption{\label{fig:kink_t} Mode evolution of the internal kink instability in
  tokamak geometry. The pullback-scheme simulation on the left is compared to
  the control-variate simulation with the same set of numerical parameters. Note
  the different length of the simulations. One sees that the control-variate
  simulation becomes numerically unstable within a few time steps.}
\end{center}
\end{figure}

\begin{figure}[h!]
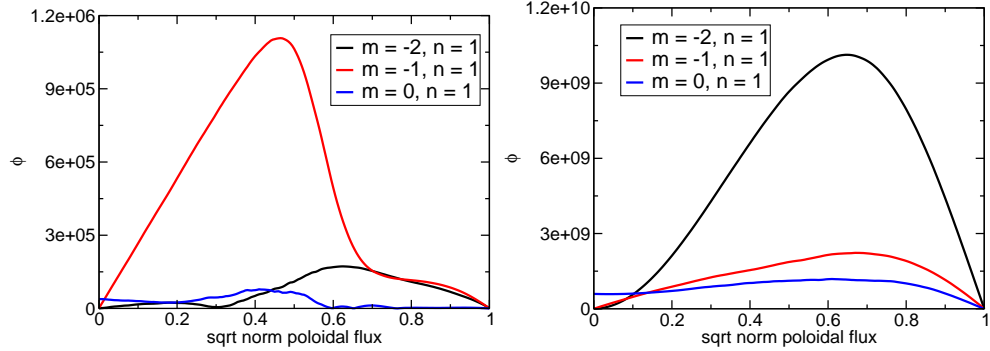

\begin{center}
\vspace*{0.5cm}
\includegraphics[width=0.47\textwidth]{kinkPB_phi_sm.eps}
\includegraphics[width=0.47\textwidth]{kinkCV_phi_sm.eps}
\caption{\label{fig:kink_sm} Mode structure of the internal kink instability in
  tokamak geometry. On the left, the radial structure of the internal
  kink mode electrostatic potential resulting from the pullback
  scheme. On the right, the mode structure obtained using the identical numerical parameters
  (number of the markers, time step, grid resolution, etc) but with only the
  control variate implemented. This mode structure is unphysical.}
\end{center}
\end{figure}
In this Section, we consider an internal kink instability in tokamak geometry. This mode is
driven by a combination of ambient parallel current and pressure
gradient. In the gyrokinetic model of ORB5, the ambient current must be
included as a modification of the background electron distribution
function, for example using a shifted Maxwellian:
\be
F_{0\e} = n_0 \left(\frac{m_{\rm e}}{2 \pi T_{\e}}\right)^{3/2} \,
\exp\left(\,-\,\frac{m_\e \veps}{T_\e}\right) \,
\exp\left[\,-\,\frac{m_\e u_0 (u_0 - 2 v_{\|})}{2 T_\e}\right]
\ee
written as a function of the energy $\veps$, the parallel velocity $v_{\|}$,
the poloidal flux $\psi$, and the poloidal angle $\theta$:
\ba
&&{} \veps = \frac{v_{\|}^2}{2} + \mu B \ ,\;\;
\mu = \frac{v_{\perp}^2}{2 B} \ ,\;\;
u_0 = \frac{j_{\|0}}{q_{\rm e} n_0}   \\
&&{} j_{\|0}(\psi, \theta) = \frac{1}{\mu_0} \vc{b}_0(\psi,\theta) \cdot \nabla
\times \vc{B}_0(\psi,\theta)
\ea
In this paper, we simulate the internal kink instability in the large aspect
ratio tokamak geometry and use $j_{\|0}(\psi,\theta=0)$ for simplicity, so that
$u_0 = u_0(\psi)$.
The spatial derivative is taken at constant $\veps$ and $v_{\|}$:
\ba
&&{} \pard{F_{0{\rm e}}}{s} = F_{0\e} \left[\frac{n'_0}{n_0} -
  \left(\frac{3}{2} - \frac{m_\e\veps}{T_\e} -
\frac{m_\e u_0 (u_0 - 2 v_{\|})}{2 T_\e}\right) \frac{T'_\e}{T_\e} -
\frac{m_\e (u_0 - v_{\|})}{T_\e} u'_0\right] \nonumber \\
&&{} n'_0 = \totd{n_0}{\psi} \ , \;\;
T'_\e = \totd{T_\e}{\psi} \ , \;\;
u'_0 = \totd{u_0}{\psi}
\ea
The parallel-velocity derivative is taken at constant $\vc{R}$ and $\veps$;
the energy derivative is taken at constant $\vc{R}$ and $v_{\|}$:
\be
\pard{F_{0\e}}{v_{\|}} = \frac{m_\e u_0}{T_\e} F_{0\e} \ ; \;\;
\pard{F_{0\e}}{m_\e \veps} = \,-\,\frac{m_\e}{m_\e T_\e} F_{0\e}
\ee
We consider a tokamak with the minor radius $r_{\rm a} = 1$~{\rm m}, the major
radius $R_0 = 10$~{\rm m}, the magnetic field at the axis $B_0 = 1$~{\rm T}, flat ion and
electron temperatures $T_\e = T_\im$ defined by $L_{x} = 2 r_{\rm a} /
\rho_{\rm s} = 360$ with $\rho_{\rm s} = \sqrt{m_\im T_\e}/(e B)$ the
sound gyroradius. The safety factor is $q(s) = 0.8 (1 + s^2)$, the flux
surface label $s = \sqrt{\psi/\psi_{\rm a}}$ with $\psi_{\rm a}$  the poloidal flux at the
plasma edge. The ambient plasma density profile $n_{0{\rm i}}(s) = n_{0{\rm e}}(s)$ is given by
\be
n_{0\e}(s) = n_0 \exp\left[\,-\,\Delta_{\rm n} \kappa_{\rm n}
  \tanh\left(\frac{s-s_0}{\Delta_{\rm n}}\right)\right]
\ee
with $\kappa_{\rm n} = 3.0$, $s_0 = 0.5$, $\Delta_{\rm n} = 0.2$, and $n_0$ corresponding
to $\beta_{\rm e} = \mu_0 n_{0\rm e} T_{\rm e}/B^2 = 0.0052$.

The mode evolution is shown in Fig.~\ref{fig:kink_t}. Here, the simulation
using the pullback mitigation is compared to the simulation applying only the control
variate (without the pullback mitigation).
In both simulations, we use $N_{\rm e} = 64 \times 10^6$ electron markers,
$N_{\rm i}
= 16 \times 10^6$ ion markers, $N_s = 200$ radial grid points, $N_{\theta} =
16$ poloidal grid points, and $N_{\varphi} = 8$ toroidal grid points. The time
step is $\omega_{c\rm i} \Delta t = 10$. The Fourier filter \cite{mcmillan_2008}
includes the poloidal modes $m \in [-2, 2]$ and the toroidal mode $n = 1$
(recall that we consider a large-aspect-ratio tokamak). The ion gyro-average
is computed using the adaptive scheme \cite{Hatzky02}, the electrons are
drift-kinetic. We initialise our simulations using an initial perturbed
distribution function with the poloidal and toroidal mode numbers $m = -1$ and
$n = 1$.

In Fig.~\ref{fig:kink_t} on the left, one sees that the pullback simulation
shows a decay of the shear Alfv\'en wave continuum and results in a
physically-driven internal kink mode developing after the decay phase is
completed. In contrast, the simulation using exclusively the control variate
becomes immediately numerically unstable. Note that the electrostatic potential
reaches very high values within just a few time steps indicating a strong
numerical instability at action. Such a pronounced instability is typical for a
too large time step.
The radial structure of the electrostatic potential developing at the end of
each simulation is shown in Fig.~\ref{fig:kink_sm}. Here again, one
sees a typical internal kink mode structure resulting from the pullback simulation
whereas the control-variate mode structure is unphysical. Further convergence
tests with a smaller time step confirm that the control variate scheme becomes
numerically stable for a time step of $\omega_{\rm ci} \Delta t = 1$.

This example shows again that the pullback approach is numerically more robust
in terms of the time step compared to the control-variate mitigation. This
observation for the kink instability is consistent with the TAE case, shown in
Figs.~\ref{fig:tae_ami} and \ref{fig:omega_mat}, where the pullback mitigation
has performed better than the control variate at larger time steps.
%
%~~~~~~~~~~~~~~~~~~~~~~~~~~~~~~~~~~~~~~~~~~~~~~~~~~~~~~~~~~~~~~~~~
%
\subsection{Nonlinear simulations} \label{nlin_pullback}
\begin{figure}[h!]
\begin{center}
\includegraphics[width=0.47\textwidth]{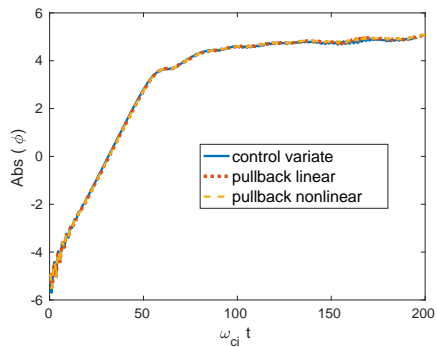}
\caption{\label{fig:nonlin} Nonlinear simulations with drift-kinetic fast
  ions. Control-variate mitigation is compared with the linear and nonlinear
  pullback schemes, see Eqs.~(\ref{ic_f}) and (\ref{ic_v}) respectively. Note
  that, not only the linear growth rate, but also the nonlinear saturation
  level agrees well with and without the pullback scheme.}
\end{center}
\end{figure}
Finally, we compare the pullback scheme with the usual control variate
approach in the nonlinear regime. In this case, one should take into account
additional nonlinear terms in the mixed-variable equations of motion
\cite{Kleiber_pullback}. The dominant nonlinear contribution is related to 
\be
\vc{B}^* = \vc{B} + \frac{m v_{\|}}{q} \nabla\times\vc{b} + \nabla
\Bgav{A_{\|}^{\rm(s)}} \times \vc{b} 
\ee
leading to 
\be
\vc{b}^* = \frac{\vc{B}^*}{B_{\|}^*} = \vc{b}^*_0 
+ \frac{\nabla \Bgav{A_{\|}^{\rm(s)}} \times \vc{b}}{B_{\|}^*} \ , \;\;
\vc{b}^*_0 \approx \vc{b} + \frac{m v_{\|}}{q B_{\|}^*} \nabla\times\vc{b}
\ee
The mixed-variable perturbed equations of motion including these nonlinear terms
are
\ba
\label{dotRnl}
\dot{\vc{R}}^{(1)} &=& \frac{\vc{b}}{B_{\|}^*} \times \nabla \Bgav{ \phi - 
  v_{\|} A_{\|}^{\rm(s)} - v_{\|} A_{\|}^{\rm(h)} } \\
&&{} - \frac{q}{m} \,\gav{A^{\rm(h)}_{\|}} \, \left( \vc{b}^*_0 + 
\frac{\nabla \Bgav{A_{\|}^{\rm(s)}} \times \vc{b}}{B_{\|}^*} \right) \nonumber\\
\label{dotvnl}
\dot{v}_{\|}^{(1)} &=& \,-\, \frac{q}{m} \, 
\left[ \vc{b}^* \cdot \nabla \Bgav{\phi - v_{\|} A_{\|}^{\rm(h)}} + \pard{}{t}
  \Bgav{A_{\|}^{\rm(s)}} \right] \\
&&{} -  \mu \, \frac{\vc{b} \times \nabla B}{B_{\|}^*} \cdot \nabla
  \Bgav{A_{\|}^{\rm(s)}} \nonumber
\ea
Substituting Ohm's law, Eq.~(\ref{Ohm}), we obtain for the parallel velocity
\ba
\dot{v}_{\|}^{(1)} &=& \,-\,  
\left( \frac{v_{\|}}{B_{\|}^*} \nabla\times\vc{b} + \frac{q}{m} \frac{\nabla
      \Bgav{A_{\|}^{\rm(s)}} \times \vc{b}}{B_{\|}^*} \right) 
\cdot \nabla \Bgav{\phi - v_{\|} A_{\|}^{\rm(h)}} 
\,+\, \\
&&{} \,+\, \frac{q}{m} \, v_{\|} \vc{b} \cdot \nabla \Bgav{A_{\|}^{\rm(h)}}  
-  \mu \, \frac{\vc{b} \times \nabla B}{B_{\|}^*} \cdot \nabla
  \Bgav{A_{\|}^{\rm(s)}} \nonumber 
\ea
For the perturbed energy, we can write 
\ba
\label{dotepsnl}
&&{} \dot{\veps}^{(1)} = v_{\|} \dot{v}_{\|}^{(1)} + \mu \dot{\vc{R}}^{(1)}
\cdot \nabla B  \ , \;\; \veps = \frac{v_{\|}^2}{2} + \mu B \ , \;\; 
\mu = \frac{v_{\perp}^2}{2 B}  \\
&&{} \dot{\veps}^{(1)} = \,-\, \frac{q_s}{m_s} \left[ m_s \mu \frac{\vc{b}
    \times \nabla B}{q_s B_{\|}^*} +
  \frac{m_s v_{\|}^2}{q_s B_{\|}^*} \, (\nabla \times \vc{b}) 
+ v_{\|} \, \frac{\nabla \Bgav{A_{\|}^{\rm(s)}} \times \vc{b}}{B_{\|}^*} \right] 
\cdot \nabla \gav{\phi}  \,+\, \nonumber \\
&&{} \,+\, \frac{q_s}{m_s} v_{\|} \left[
   v_{\|}\vc{b} + m_s \mu \frac{\vc{b} \times \nabla B}{q_s B_{\|}^*} +
 \frac{m_s v_{\|}^2}{q_s B_{\|}^*} \, (\nabla \times \vc{b}) + 
v_{\|} \, \frac{\nabla \Bgav{A_{\|}^{\rm(s)}} \times \vc{b}}{B_{\|}^*} \right] 
\cdot \nabla \gav{A_{\|}^{\rm(h)}} \,+\, \nonumber \\
&&{} \,+\, \frac{q_s}{m_s} \mu B \left[ \nabla\cdot\vc{b} - \frac{m v_{\|}}{q B_{\|}^*}
  \frac{\nabla \times \vc{B}}{B^2} \cdot \nabla B - 
\frac{\nabla \Bgav{A_{\|}^{\rm(s)}}}{B_{\|}^*} \cdot 
\frac{\vc{b}\times\nabla B}{B} \right] \Bgav{A_{\|}^{\rm(h)}}
\ea
Note that the transformation Eq.~(\ref{ic_f}) is perturbative in the field
$A_{\|}^{\rm(h)}$. To make it fully nonlinear, we should modify the pullback
algorithm to the following \cite{Kleiber_pullback}. 
\begin{enumerate}
\item At the end of each time step, redefine the magnetic potential splitting,
  collecting the entire instantaneous value of $A_{\|}(t_i)$ in its `symplectic part': 
\be
A_{\|\rm(new)}^{\rm(s)}(t_i) = A_{\|}(t_i) = A_{\|\rm(old)}^{\rm(s)}(t_i) +
A_{\|\rm(old)}^{\rm(h)}(t_i) 
%\longrightarrow %\ , \;\;\; A_{\|}^{\rm(h)}(t_i) \longrightarrow 0
\ee
\item As a consequence of the new splitting, Eq.~(\ref{ic_A}), the
  `hamiltonian' part of the vector potential must be corrected:
\be
A_{\|\rm(new)}^{\rm(h)}(t_i) = 0
\ee
\item Transform the phase-space coordinates keeping the particle weights
  constant (this part is modified compared to Sec.~\ref{theory}):
\ba
\label{ic_v}
&&{} v_{\|\rm(new)}^{\rm(m)} = v_{\|}^{\rm(s)} = 
v_{\|\rm(old)}^{\rm(m)} - \frac{q_s}{m_s} \Bgav{A_{\|}^{\rm(h)}(t_i)} \\
\label{ic_f1}
&&{} f_{1s\rm(new)}^{\rm(m)}\left(v_{\|\rm(new)}^{\rm(m)}\right) = %f_{1s}^{\rm(s)}(v_{\|\rm(new)}^{\rm(m)}) = 
f_{1s\rm(old)}^{\rm(m)}\left(v_{\|\rm(old)}^{\rm(m)}\right)
\ea
\item Proceed, explicitly solving the mixed-variable system of equations 
  at the next time step $t_i + \Delta t$ in a usual way, but using
  Eqs.~(\ref{ic_v}) and (\ref{ic_f1}) as the initial conditions. 
%
%\item At the end of the time step, repeat the transformation
%  Eqs.~(\ref{ic_A})-(\ref{Ah_new}). 
\end{enumerate}

In Fig.~\ref{fig:nonlin}, we consider the nonlinear TAE mode \cite{MCole_PoP16}
with only the wave-particle nonlinearity kept comparing the three mitigation
schemes: the control variate, the linear pullback Eq.~(\ref{ic_f}), and the
nonlinear pullback, Eqs.~(\ref{ic_v}) and (\ref{ic_f1}), with the nonlinear
terms included in the equations of motion, see Eqs.~(\ref{dotRnl}), (\ref{dotvnl}) and
(\ref{dotepsnl}). 
Note that in the case of ``linear pullback'' we use the equations of motion
(\ref{dotR1}) and (\ref{dotp1}). These equations include the perturbed fields, which
makes the simulation itself nonlinear, but we ignore the higher-order
nonlinear terms that are present in Eqs.~(\ref{dotRnl}) and (\ref{dotvnl}).
%
%Note that in the case of ``linear pullback'' we still
%follow the perturbed fast-particle orbits, which makes the simulation itself
%nonlinear, but we ignore the nonlinear terms in the equations of motion and
Also, we use the transformation Eq.~(\ref{ic_f}) at the end of every time
step. In the case of ``nonlinear pullback'' all the terms are included in the
equations of motion and the nonlinear transformation, Eqs.~(\ref{ic_v}) and
(\ref{ic_f1}), is employed. 

In our simulations, we use $N_{\im} = 10^7$ ion markers, $N_\e =
4\times10^7$ electron markers, $N_{\rm f} = 10^8$ nonlinear fast-ion markers,
$256$ grid points in the radial and poloidal directions, and $64$ grid points in
the toroidal direction. The time step is $\omega_{{\rm ci}} \Delta t = 20$
for the control variate scheme and $\omega_{{\rm ci}} \Delta t = 100$ for the
pullback mitigation. We use a reduced mass ratio $m_e / m_i = 0.005$
and drift-kinetic fast ions. 
%
%In Fig.~\ref{fig:nonlin}, 
One sees that both the saturation levels and the linear evolution agree very
well in Fig.~\ref{fig:nonlin} for all the schemes. This verifies the new
pullback implementation in ORB5 also nonlinearly. 
Interestingly, the discrepancy between the ``linear pullback'' and ``nonlinear
pullback'' in terms of the saturation level was larger for the tearing mode
\cite{Kleiber_pullback} where the electron nonlinearity was important. Here,
we see that this difference is negligible for the fast-particle
nonlinearity. Thus, the original pullback scheme and the mixed-variable
equations described in Sec.~\ref{theory} can safely be used for the Alfv\'enic
simulations where the fast-ion nonlinearity is dominant. 
%This case has been considered in more detail using only the
%control-variate mitigation in Ref.~\cite{MCole_PoP16}. 
%
%=================================================================
%
\section{Conclusions}\label{conclusions}
The pullback scheme has been implemented in ORB5. The solver modifications needed
for the pullback scheme implementation have been described. The new scheme has
been verified using the ITPA-TAE benchmark \cite{Koenies_itpa,Koenies_itpa_NF}
both in the linear and nonlinear regimes. A considerable improvement of the
code efficiency has been observed. Also, the efficiency of the pullback
mitigation in ORB5 has been demonstrated using the internal kink mode in
tokamak geometry. To our knowledge, internal kink mode simulations in
tokamak geometry have not been reported previously for global gyrokinetic
particle-in-cell codes, such as ORB5, using realistic values for the plasma~$\beta$.

In the outlook, ORB5 provides a unified framework which includes
electromagnetic drift-wave turbulence, zonal flows and GAMs, fast particles,
shear Alfv\'en waves (TAEs, BAEs, etc), and MHD activity (tearing mode,
internal kink instability) in axisymmetric tokamak geometry. This represents a
vast field for future research.
%
%%%%%%%%%%%%%%%%%%%%%%%%%%%%%%%%%%%%%%%%%%%%%%%%%%%%%%%%%%%%%%%%%

\qquad \\
\qquad \\
{\bf Acknowledgements}
%%%%%%%%%%%%%%%%%%%%%%%%%%%%%%%%%%%%%%%%%%%%%%%%%%%%%%%%%%%%%%%%%
%
We acknowledge P.~Helander, E.~Sonnendr\"ucker, F.~Jenko, S.~G\"unter,
F.~Zonca, and Ph.~Lauber for their support. Numerical
simulations were performed on the Marconi supercomputer within the framework
of the OrbZONE and OrbFAST projects. This work has been carried out within the
framework of the EUROfusion Consortium and has received funding from the
Euratom research and training program 2014--2018 under Grant Agreement
No. 633053, for the CfP-AWP17-ENR-MPG-01 (2017/2018) project on ``Nonlinear
interaction of Alfv\'enic and turbulent fluctuations in burning plasmas
(NAT)''. The views and opinions expressed herein do not necessarily reflect
those of the European Commission.
%
%=================================================================
%
%%%%%%%%%%%%%%%%%%%%%%%%%%%%%%%%%%%%%%%%%%%%%%%%%%%%%%%%%%%%%%%%%
% \appendix
% \section{}\label{appA}
% This appendix contains sample equations in the JPP style. Please refer to the
% {\LaTeX} source file for examples of how to display such equations in your
% manuscript.

%%%%%%%%%%%%%%%%%%%%%%%%%%%%%%%%%%%%%%%%%%%%%%%%%%%%%%%%%%%%%%%%%
% susie put cite commands here, don't bother with citet etc just yet.
% Note the spaces between the initials

%\bibstyle{prsty}

\bibliographystyle{model1-num-names}
%\bibliography{\BIBL/gyrokin,\BIBL/gygles,\BIBL/aitg}

%-------------------------------------------------
%
%\graph{hsx}{Magnetic field configuration of the HSX stellarator \cite{Almagri1999}.}
%
%-------------------------------------------------

\end{document}